\documentclass[pra,twocolumn,showpacs]{revtex4}
\usepackage{amsmath}
\usepackage{amssymb}
\usepackage{epsfig}

\newcommand{\omegabold}{\boldsymbol\omega}

\begin{document}

\title{\bf The dynamics of straight vortex filaments in a Bose-Einstein condensate with a Gaussian density profile}
\author{V.~P. Ruban}
\email{ruban@itp.ac.ru}
\affiliation{L.D. Landau Institute for Theoretical Physics RAS, Moscow, Russia} 

\date{\today}

\begin{abstract}
The dynamics of interacting quantized vortex filaments in a rotating trapped Bose-Einstein condensate,
which is in the Thomas-Fermi regime at zero temperature and described by the Gross-Pitaevskii equation,
is considered in the hydrodynamic ``anelastic'' approximation. In the presence of a smoothly inhomogeneous
array of filaments (vortex lattice), a non-canonical Hamiltonian equation of motion is derived for the 
macroscopically averaged vorticity, with taking into account the spatial non-uniformity of the equilibrium
condensate density determined by the trap potential. A minimum of the corresponding Hamiltonian describes
a static configuration of deformed vortex lattice against a given density background. The minimum condition
is reduced to a vector nonlinear partial differential equation of the second order, for which some
approximate and exact solutions are found. It is shown that if the condensate density has an anisotropic
Gaussian profile then equation of motion for the averaged vorticity admits solutions in the form of a 
spatially uniform vector with a nontrivial time dependence. An integral representation is obtained for 
the matrix Green function determining the non-local Hamiltonian of a system of arbitrary shaped
vortex filaments in a condensate with Gaussian density. In particular, if all the filaments 
are straight and parallel to one of the main axes of the ellipsoid, then a finite-dimensional reduction follows 
which is able to describe the dynamics of a system of point vortices at a non-uniform background.
A simple approximate expression for the two-dimensional Green function is suggested at rather arbitrary
density profile, and its successful comparison to the exact result in the Gaussian case is done. 
Approximate equations of motion are derived which describe a long-wave dynamics of interacting vortex 
filaments in condensates with the density depending on the transverse coordinates only.
\end{abstract}
\maketitle

\section{Introduction}

For theoretical study of the dynamics of quantized vortex filaments in a rotating trapped Bose-Einstein condensate, 
which is in the Thomas-Fermi regime at zero temperature and described by the Gross-Pitaevskii equation,
the hydrodynamic approximation is often used (see, e.g., \cite{SF2000,FS2001,F2009,A2002,SR2004}, 
and many references therein). In a coordinate system rotating  with an angular velocity ${\bf\Omega}$ together
with the trap potential $V({\bf r})$, the behaviour of a relatively thin vortex filament with a good accuracy
obeys classical equations of slow hydrodynamics of an inviscid compressible fluid against a given
spatially non-uniform static density background $\rho_0({\bf r})\approx\mbox{const}\cdot[\mu- V({\bf r})]$, 
where $\mu$ is the chemical potential of the bosons. In other words, the condensate is treated in
``anelastic'' approximation, i.e., instead of the full continuity equation, the condition 
$\nabla\cdot (\rho_0{\bf v})=0$  is imposed, thereby excluding from the consideration all potential 
excitations, in particular surface modes. It  is implied herewith that $\Omega\ll\omega_\perp$,  where $\omega_\perp$
is a characteristic transverse frequency of the trap. First, this inequality ensures the condensate stability
\cite{RZS2001,SC2001}. Second, since a centrifugal-force-caused additional equilibrium density profile
deformation has the relative order $(\Omega/\omega_\perp)^2$, it is possible to neglect this effect.
At the same time, $\mu$ is sufficiently large,  $[\mu-V_{\rm min}]\gg\hbar\omega_\perp$, and it is easy to  check
this inequality results in small relation of a vortex core width $\xi({\bf r})\propto\rho_0^{-1/2}$ 
to  a characteristic transverse condensate size $R_\perp$ almost everywhere in the bulk. The condition
$$R_\perp/\xi \sim ([\mu-V_{\rm min}]/\hbar\omega_\perp)[\rho_0({\bf r})/\rho_0(0)]^{1/2}\gg 1$$
is only broken near the Thomas-Fermi surface $[\mu-V({\bf r})]=0$.
The spatial inhomogeneity of the density contributes essentially to the statics and dynamics of vortices
\cite{SF2000,FS2001,F2009,A2002,SR2004,AR2001,GP2001,RBD2002,AD2003,AD2004,D2005,Kelvin_vaves,ring_istability}, 
but this question is still far from being completely investigated. In this work, we will present  some new results 
obtained in this direction with the help of the Hamiltonian formalism.

Before formulating the purposes of the work, it is necessary to remind briefly some basic facts
about Hamiltonian mechanics of vortex filaments. It is well known that in classical hydrodynamics
the continuous vorticity field $\omegabold\equiv\nabla\times{\bf v}_{\rm lab}$ in the laboratory frame of 
reference obeys the equation
\begin{equation}\label{omega_lab}
\omegabold_t=\mbox{curl}[{\bf v}_{\rm lab}\times\omegabold].
\end{equation}
The velocity field  ${\bf v}_{\rm lab}=\nabla\varphi_0+{\bf v}$ consists of two terms. First, it is
an explicitly time-depending  potential velocity field  $\nabla\varphi_0({\bf r},t)$ of the liquid medium
in the absence of vortices, which field is created by a transverse anisotropy of the rotating trap potential.
When we use the rotating frame of reference, then $\nabla\varphi_0$ ceases to depend on time, but additional
term $-[{\bf\Omega}\times{\bf r}]$ appears, so the unperturbed fluid velocity becomes 
${\bf v}_0=\nabla\varphi_0-[{\bf\Omega}\times{\bf r}]$. In anelastic approximation the condition
$\nabla\cdot (\rho_0{\bf v}_0)=0$ is satisfied, which means the existence of a vector potential 
${\bf A}_0({\bf r})$:
\begin{equation}\label{v_0}
\rho_0{\bf v}_0=\mbox{curl}\,{\bf A}_0, \qquad \mbox{curl}\,{\bf v}_0=-2{\bf\Omega}.
\end{equation}
Second,  ${\bf v}_{\rm lab}$ has the part ${\bf v}$ created by vorticity, i.e.
\begin{equation}\label{v_cond}
\rho_0{\bf v}=\mbox{curl}\,{\bf A}, \qquad \mbox{curl}\,{\bf v}={\omegabold}.
\end{equation}
In other words, the vector potential  ${\bf A}$ of the mass current density ${\bf j}=\rho_0{\bf v}$ 
has to be found from the equation
\begin{equation}\label{A_eq}
\mbox{curl}\frac{1}{\rho_0({\bf r})}\mbox{curl}{\bf A}={\omegabold}({\bf r}).
\end{equation}
A general solution of Eq.(\ref{A_eq}) is expressed through a (matrix) Green's function 
$\hat G({\bf r}_2,{\bf r}_1)$, which is symmetric on the arguments:
\begin{equation}
 {\bf A}({\bf r})=\int \hat G({\bf r},{\bf r}_1){\omegabold}({\bf r}_1)d {\bf r}_1.
\end{equation}
It is easy to see that 
\begin{equation}
 {\bf A}_0({\bf r})=-2\int \hat G({\bf r},{\bf r}_1){\bf\Omega}\,d {\bf r}_1.
\end{equation}

In the rotating coordinate system, instead of Eq.(\ref{omega_lab}) we have
\begin{equation}\label{omega_rot}
\omegabold_t=\mbox{curl}[({\bf v}_0+{\bf v})\times\omegabold].
\end{equation} 
In fact, equations (\ref{v_0}), (\ref{v_cond}), and (\ref{omega_rot}) determine a non-canonical
Hamiltonian system
\begin{equation}\label{non_canon}
{\omegabold}_t=\mbox{curl}\Big[\frac{1}{\rho_0}\mbox{curl}\Big(\frac{\delta{\cal H}}
{\delta{\omegabold}}\Big)\times{\omegabold}\Big]
\end{equation}
with the Hamiltonian
\begin{eqnarray}\label{Hamiltonian}
{\cal H}\{{\omegabold}\}&=&\frac{1}{2}\int\!\!\int {\omegabold}({\bf r}_2)\cdot 
\hat G({\bf r}_2,{\bf r}_1){\omegabold}({\bf r}_1)d {\bf r}_1 d {\bf r}_2\nonumber\\
&+&\int {\bf A}_0({\bf r})\cdot{\omegabold}({\bf r})d {\bf r}.
\end{eqnarray} 
Let us note that the double integral here is the kinetic energy of a vortical
flow in the absence of rotation.

We consider vorticity distributions in the form of relatively narrow vortex tubes imitating singular
quantum vortex filaments with the velocity circulation $\Gamma=2\pi\hslash/m_{\rm atom}$. To investigate
such configurations, it is convenient to use the so called vortex line representation 
(see details in  \cite{KR1998,R2001,R2003}),
\begin{eqnarray}\label{VLR}
\omegabold({\bf r},t)=\int_{\cal N}d^2\nu\oint\delta({\bf r}-{\bf R}(\nu,\beta,t)){\bf R}_\beta d\beta,
\end{eqnarray}
where $\nu=(\nu_1,\nu_2)\in{\cal N}$ is a vortex line label situated in a two-dimensional (2D) manifold
${\cal N}$, and  $\beta$ is an arbitrary longitudinal parameter along the line.
Formula (\ref{VLR}) represents a divergence-free vorticity field as a distribution of closed vortex lines
in the three-dimensional (3D) space. Such a parametrization allows us to obtain a variational equation of motion 
for a vortex line from the non-canonical equation (\ref{non_canon}),
\begin{equation}\label{var_eq_nu}
[{\bf R}_\beta\times{\bf R}_t]\rho_0({\bf R})=\delta {\cal H}_{\rm VL}/\delta{\bf R}(\nu,\beta),
\end{equation}
where the Hamiltonian of vortex lines ${\cal H}_{\rm VL}\{{\bf R}(\nu,\beta)\}$ is obtained by substitution
of expression (\ref{VLR}) into the Hamiltonian (\ref{Hamiltonian}). In the limit of a single infinitely narrow
vortex filament, the vector function  ${\bf R}(\nu,\beta,t)$ ceases to depend on $\nu$, and the velocity 
circulation is  $\Gamma=\int_{\cal N}d^2\nu$. At that, the equation of motion 
of a thin vortex filament follows from a variational principle with the Lagrangian of the form
\cite{R2001,R2003,R2016}
\begin{equation}\label{L_filament}
{\cal L}=\Gamma \oint({\bf D}({\bf R})\cdot[{\bf R}' \times{\bf R}_t]) d\beta -{\cal H}\{{\bf R}\},
\end{equation}
where ${\bf R}'\equiv {\bf R}_\beta(\beta, t)$ is the tangent vector, while the vector function 
${\bf D}({\bf r})$ satisfies the condition 
\begin{equation}
\nabla\cdot{\bf D}({\bf r})=\rho_0({\bf r}).
\end{equation}
The corresponding variational equation of motion has the structure
\begin{equation}\label{var_eq}
\Gamma[{\bf R}'\times{\bf R}_t]\rho_0({\bf R})=\delta {\cal H}/\delta{\bf R}(\beta),
\end{equation}
with the Hamiltonian
\begin{equation}\label{H_discr}
{\cal H}\{{\bf R}\}={\cal K}_\Gamma\{{\bf R}\} +\Gamma\oint ({\bf A}_0({\bf R})\cdot {\bf R}') d \beta.
\end{equation}
Herewith, the kinetic energy of the vortex  ${\cal K}_\Gamma\{{\bf R}\}$ is given by the formula
\begin{equation}\label{K}
{\cal K}_\Gamma=\frac{\Gamma^2}{2}\oint\!\!
\oint {\bf R}'(\beta_2)\cdot \hat G({\bf R}(\beta_2),{\bf R}(\beta_1))
{\bf R}'(\beta_1)d\beta_1 d\beta_2.
\end{equation}
Since at ${\bf r}_1\to{\bf r}$ the asymptotics 
\begin{equation}\label{G_asympt} 
\hat G({\bf r},{\bf r}_1)\approx \rho_0({\bf r})/(4\pi|{\bf r}-{\bf r}_1|)
\end{equation}
takes place, the given logarithmically divergent integral is cut at a core width $\xi$.
At this point of the theory, some difference appears between the classical anelastic hydrodynamics where
the diameter of a thin vortex filament at any time moment is almost constant along the central line, and  
Bose-Einstein condensates where $\xi$ depends on ${\bf r}$ through the equilibrium density.
However, under condition $\log(R_\perp/\xi)\gg 1$ this difference is not very important for the vortex dynamics.

Generalization to the case of several vortex lines is evident --- each integration over the longitudinal
parameter should be supplemented with summation over all the vortices.

A technical difficulty arises since for most cases of physically relevant density profiles it is impossible
to calculate the Green function analytically. Therefore in the general case one cannot go beyond the so called
local induction approximation (LIA)  \cite{SF2000,R2001,R2016}, which is based on the approximate equality
(\ref{G_asympt}):
\begin{equation}\label{H_LIA}
{\cal K}_\Gamma\{{\bf R}\}\approx({\Gamma^2\Lambda}/{4\pi})\oint \rho_0({\bf R}) |{\bf R}_\beta|d\beta,
\end{equation}
where $\Lambda=\log(R_\perp/\xi)\approx\mbox{const}\approx\log([\mu-V_{\rm min}]/\hslash\omega_\perp) \gg 1$
is the large logarithm. The equation of motion (\ref{var_eq}), corresponding to this functional, is
\begin{eqnarray}
&&\Gamma[{\bf R}_\beta\times{\bf R}_t]\rho_0({\bf R})=\Gamma[{\bf R}_\beta\times\mbox{curl}{\bf A}_0({\bf R})]
\nonumber\\
&&\quad+\frac{\Gamma^2\Lambda}{4\pi}\Big\{\nabla\rho_0({\bf R})|{\bf R}_\beta|
-\frac{\partial}{\partial\beta}\Big(\rho_0({\bf R})\frac{{\bf R}_\beta}{|{\bf R}_\beta|}\Big)\Big\}.
\label{LIA_variat}
\end{eqnarray}
After resolution with respect to the time derivative, it takes the form
\begin{equation}\label{LIA}
{\bf R}_t =\frac{\Gamma\Lambda}{4\pi} \Big\{\varkappa {\bf b} 
+\Big[\frac{\nabla \rho_0({\bf R})}{\rho_0({\bf R})}\times \frac{{\bf R}_\beta}{|{\bf R}_\beta|}\Big]
\Big\} +{\bf v}_0({\bf R}),
\end{equation}
where $\varkappa$ is the local curvature of the line,  ${\bf b}$ is the unit bi-normal vector, and
${\bf R}_\beta/|{\bf R}_\beta|$ is the unit tangent vector. Clearly, such an approximation is valid for a single
vortex only, and it is not able to describe the interaction of several filaments.

In this work, first, a non-canonical Hamiltonian system of equations for a macroscopically smoothed vorticity
will be derived, in the presence of many quantized vortex filaments on a spatially inhomogeneous density background.
Second, we will investigate one of the few exclusive cases when an analytical expression for $\hat G$  still
can be presented. Namely, a Gaussian profile of the equilibrium density inside the condensate 
(i.e., not too closely to the Thomas-Fermi surface) is meant:
$\rho_0({\bf r})=\exp(-{\bf r}\cdot \hat S {\bf r})$, where $\hat S$ is a symmetric positively posed matrix 
$3\times 3$ (for simplicity of calculations, but without loss of generality --- diagonal one, with eigenvalues
 $\{s_1,s_2,s_3\}$). Such a dependence qualitatively resembles an inverted paraboloid
$[1-{\bf r}\cdot \hat S {\bf r}]$ taking place in most simple --- harmonic --- traps. 
Besides that, a close-to-Gaussian density can be prepared in anharmonic potential wells corresponding to expansion
of the exponent on the powers of its argument up to an odd order, for example,
$(\mu-V)\propto 1-({\bf r}\cdot \hat S {\bf r}) + ({\bf r}\cdot \hat S {\bf r})^2/2 
-({\bf r}\cdot \hat S {\bf r})^3/6$.

Two interesting facts deserve mentioning here, which are characteristic namely for Gaussian density profiles
in the context of vortex dynamics. First, the local induction equation (\ref{LIA}) on such backgrounds admits
nontrivial solutions in the form of a straight nonstationary vortex \cite{R2016-2}. Second, as it will be shown in
Section II, in the quasi-continuous limit there are also exact nonstationary solutions, in this case corresponding
to spatially uniform field of the effective vorticity which obeys an integrable ordinary differential equation.

The presence of the indicated integrable reductions, as well as results of work \cite{SR2004} where the influence
of density inhomogeneity upon properties of vortex lattices in 2D systems was studied and it was found 
that at Gaussian background the lattice remains uniform, suggest an idea that Gaussian density profile
takes a special place among others, as far as the dynamics of quantum vortex filaments is concerned.
Therefore it is reasonable to study properties of such systems in more detail. In particular, in Section III the 
corresponding Green function will be calculated, and in Section IV a finite-dimensional reduction will be
investigated, which describes the motion of several filaments parallel to one of the main axes of the ellipsoid
${\bf r}\cdot \hat S {\bf r}=$ const.

\section{Macroscopically averaged system}

So, we consider now the quasi-continuous limit. Let in the condensate be present a big number $N$ of mutually
unknotted vortex filaments, more-less regularly distributed in the space. In other words, locally in a cross-section
the vorticity field looks like a deformed vortex lattice without defects (dislocations, vacancies, embedded vortices
in the lattice are not considered now). Approximate macroscopic equations of motion for such a system can be obtained
through the replacement of a discrete vortex filament index $n$ by a pair of continuous quantities (Clebsch variables)
$\nu_1$ and $\nu_2$, that, roughly speaking, numerate vortex rows along two independent directions.
Discrete Lagrangian is then replaced by the continuous variant
\begin{equation}\label{L_cont}
{\cal L}_\nu=\Gamma\int ({\bf D}({\bf R})\cdot[{\bf R}' \times{\bf R}_t]) d\beta d\nu_1 d\nu_2
-{\cal H}_\nu\{{\bf R}(\nu_1,\nu_2,\beta)\},
\end{equation}
where ${\cal H}_\nu$ is an approximate Hamiltonian obtained by substitution into the kinetic energy (\ref{K}) 
[where the integration over  $\beta$ should be supplemented with summation over $n$] of the vortex array
corresponding to integer values of variables $\nu_1$ and $\nu_2$ within some 2D domain.
An accurate calculation of functional ${\cal H}_\nu\{{\bf R}(\nu_1,\nu_2,\beta)\}$ is a quite nontrivial
procedure because interactions between near neighbours are smoothed not entirely, so in the Hamiltonian
${\cal H}_\nu$ a quasi-elastic shear energy appears. As is known, such an energy makes possible specific
oscillations of the lattice in the form of Tkachenko waves
\cite{Tkach_w-1,Tkach_w-2,Tkach_w-3,Tkach_w-4,Tkach_w-5,Tkach_w-6,Tkach_w-7}. 
But if we neglect the shear energy of the lattice in comparison with a logarithmically large contribution
from the local induction, then in the main approximation ${\cal H}_\nu$ will in fact only depend upon
macroscopically averaged vorticity  $\tilde{\omegabold}({\bf r},t)$, being determined by the expression
\begin{eqnarray}\label{tilde_H}
&&\tilde{\cal H}\{\tilde{\omegabold}\}=\frac{1}{2}\int\!\!\int \tilde{\omegabold}({\bf r}_2)\cdot 
\hat G({\bf r}_2,{\bf r}_1)\tilde{\omegabold}({\bf r}_1)d {\bf r}_1 d {\bf r}_2\nonumber\\
&&\quad+\int {\bf A}_0({\bf r})\cdot\tilde{\omegabold}({\bf r})d {\bf r}+
\frac{\Gamma\tilde\Lambda}{4\pi}\int \rho_0({\bf r})|\tilde{\omegabold}({\bf r})|d {\bf r},
\end{eqnarray}
in which one needs to do replacements like $\tilde{\omegabold}({\bf r})d{\bf r}\to {\bf R}' d\beta d\nu_1 d\nu_2$.
Renormalization of the local induction coefficient,
\begin{equation}
\tilde\Lambda=\log\Big(\frac{R_\perp/\sqrt{N}}{\xi})=(\Lambda-\log\sqrt{N})\gg 1,
\end{equation}
takes into account the fact that a local contribution to the energy is accumulated on a scale as
a distance between neighbour vortices, that is $R_\perp/\sqrt{N}$, which is still supposed be large in comparison 
with a vortex core width. Thus, the given approximation does not make difference between a regular vortex lattice
and a lattice with defects. Clebsch variables in such a case lose their original meaning as numbers of lattice rows, 
and they now admit arbitrary transformations preserving an area element in the 2D space $(\nu_1,\nu_2)$.

It should be noted for completeness that in Eulerian representation the macroscopic Lagrangian for
Clebsch variables $\nu_1({\bf r},t)$ and $\nu_2({\bf r},t)$ looks as follows,
\begin{equation}\label{L_Clebsch}
\tilde{\cal L}=\Gamma\int \rho_0({\bf r}) \nu_2 \partial_t\nu_{1} d{\bf r}
-\tilde{\cal H}\{\Gamma[\nabla\nu_1\times\nabla\nu_2]\}.
\end{equation}
The corresponding equations of motion are almost canonical,
\begin{equation}\label{nu1nu2}
\partial_t\nu_{1,2}=\pm\frac{1}{\Gamma\rho_0} 
\Big(\frac{\delta\tilde{\cal H}}{\delta\nu_{2,1}}\Big)
=-\frac{1}{\rho_0}\mbox{curl}\Big(\frac{\delta\tilde{\cal H}}{\delta\tilde{\omegabold}}\Big)
\cdot\nabla\nu_{1,2},
\end{equation}
and in fact they are transport equations, with the transport velocity
\begin{equation}
{\bf V}_{\rm tr}=\frac{1}{\rho_0}\mbox{curl}\Big(\frac{\delta\tilde{\cal H}}{\delta\tilde{\omegabold}}\Big).
\end{equation}
Since the approximate Hamiltonian $\tilde{\cal H}$ only depends upon 
$\tilde{\omegabold}=\Gamma[\nabla\nu_1\times\nabla\nu_2]$, but not upon other combinations of gradients
of the Clebsch variables (which combinations inevitably would appear if the shear energy is taken into account),
the presence of this symmetry results in additional integrals of motion of the form
\begin{equation}
Q_{(\Psi)}=\int \Psi(\nu_1,\nu_2)\rho_0({\bf r}) d {\bf r},
\end{equation}
where $\Psi(\nu_1,\nu_2)$ is an arbitrary function. Herewith, the equation of motion corresponding to the
Lagrangian (\ref{L_cont}) [or to the Lagrangian (\ref{L_Clebsch})] possesses the same standard noncanonical
Hamiltonian structure of the hydrodynamic type, as equation (\ref{non_canon}):
\begin{equation}
\tilde{\omegabold}_t=\mbox{curl}\Big[\frac{1}{\rho_0}\mbox{curl}\Big(\frac{\delta\tilde{\cal H}}
{\delta \tilde{\omegabold}}\Big)
\times\tilde{\omegabold}\Big].
\end{equation}
Thus, taking into account expression (\ref{tilde_H}), we arrive at equation
\begin{equation}\label{macro}
\tilde{\omegabold}_t=\mbox{curl}\Big[\Big\{{\bf v}_0+\tilde{\bf v}+\frac{\Gamma\tilde\Lambda}{4\pi\rho_0}
\mbox{curl}\Big(\rho_0\frac{\tilde{\omegabold}}{|\tilde{\omegabold}|}\Big)\Big\}\times\tilde{\omegabold}\Big],
\end{equation}
with the conditions (\ref{v_0}) and
\begin{equation}\label{v}
\mbox{curl\,}\tilde{\bf v}=\tilde{\omegabold},\qquad \nabla\cdot(\rho_0\tilde{\bf v})=0.
\end{equation}
It is easy to see that in the ``frozen-in vorticity'' equation (\ref{macro}) the expression in curly brackets
--- transport velocity of vortex lines --- differs from the r.h.s. of equation (\ref{LIA}) for a single
filament just by renormalized $\tilde\Lambda$  and by adding a self-consistent macroscopically smoothed field
$\tilde{\bf v}$ created by all the other vortices.

Important application of this theory is description of motionless equilibrium states of a vortex lattice.
Generally speaking, for equation (\ref{macro}) it is necessary to distinguish between ``static'' and ``stationary''
solutions on which $\tilde{\omegabold}$ is independent on time. In stationary states the expression in square
brackets in equation (\ref{macro}) is a gradient of some function $\Psi(\nu_1,\nu_2)$. In static states [clearly,
they correspond to a minimum of the Hamiltonian (\ref{tilde_H})] the transport velocity of vortices is zero:
\begin{equation}
{\bf v}_0+\tilde{\bf v}_{\rm stat}+\frac{\Gamma\tilde\Lambda}{4\pi\rho_0}
\mbox{curl}\Big(\rho_0\frac{\tilde{\omegabold}_{\rm stat}}{|\tilde{\omegabold}_{\rm stat}|}\Big)=0.
\end{equation}
Acting by the curl operator, we obtain from here a closed differential equation for 
$\tilde{\omegabold}_{\rm stat}({\bf r})$:
\begin{equation}\label{stat}
\tilde{\omegabold}_{\rm stat}=2{\bf\Omega}-\frac{\Gamma\tilde\Lambda}{4\pi}
\mbox{curl}\frac{1}{\rho_0}\mbox{curl}\Big(\rho_0\frac{\tilde{\omegabold}_{\rm stat}}
{|\tilde{\omegabold}_{\rm stat}|}\Big).
\end{equation}
Thus, contrary to the dynamics, consideration of static states of vortex lattice within the given approximation
is possible without explicit formula for the Green function. To the best author's knowledge, the 3D
equation (\ref{stat}) is a new result. As far as the second term in the r.h.s. is in fact a small correction
(not too closely to condensate surface) in comparison with $2\Omega\approx N\Gamma/\pi R_\perp^2$, this equation
can be approximately solved by iterations. The reuslts of the two first iterations are given below (where the notation
${\bf e}\equiv {\bf\Omega}/|{\bf\Omega}|$ for the unit vector along the rotaion axis has been introduced):
\begin{eqnarray}
\tilde{\omegabold}^{(1)}_{\rm stat}&=& 
2{\bf\Omega}-\frac{\Gamma\tilde\Lambda}{4\pi}[({\bf e}\cdot\nabla)\nabla\log\rho_0-{\bf e}\Delta \log\rho_0],\\
\tilde{\omegabold}^{(2)}_{\rm stat}&=& 2{\bf\Omega}-\frac{\Gamma\tilde\Lambda}{4\pi}
\mbox{curl}\frac{1}{\rho_0}\mbox{curl}\Big(\rho_0\frac{\tilde{\omegabold}^{(1)}_{\rm stat}}
{|\tilde{\omegabold}^{(1)}_{\rm stat}|}\Big).
\end{eqnarray}
Besides these 3D approximate solutions, for  $\rho_0=Z(z)\rho({\bf r}_\perp)$ and rotation around
$z$ axis, there are also exact 2D solutions,
\begin{equation}\label{2D_lattice}
\tilde{\omegabold}_{\rm stat}({\bf r}_\perp)={\bf e}_z
\Big[2\Omega+\frac{\Gamma\tilde\Lambda}{4\pi}\Delta_\perp\log \rho({\bf r}_\perp)\Big].
\end{equation}
So, the main result of work \cite{SR2004}, formula (\ref{2D_lattice}), is reproduced here in a more compact manner
than in the original, moreover for arbitrary $\rho({\bf r}_\perp)$, not only for axisymmetric densities.
Let us note, in Gaussian case the macroscopic vorticity remains uniform, but slightly different from $2\Omega$.

It is also clear that for Gaussian $\rho_0({\bf r})$ only, the system (\ref{v_0}), (\ref{macro}), (\ref{v})
admits substitution $\tilde{\omegabold}={\bf M}(t)$ with an arbitrarily oriented initial vector  ${\bf M}(0)$.
At that, the velocity fields  $\tilde{\bf v}$ and ${\bf v}_0$ are both linear on ${\bf r}$:
\begin{eqnarray}
&&\tilde{\bf v}=[(\hat I \mbox{Tr}\hat S -\hat S)^{-1}{\bf M}\times \hat S{\bf r}],\\
&&{\bf v}_0=-2[(\hat I \mbox{Tr}\hat S -\hat S)^{-1}{\bf\Omega}\times \hat S{\bf r}],
\end{eqnarray}
where $\hat I$ is the unit matrix. By the way, vector potential ${\bf A}_0({\bf r})$ is determined by a
simple expression
\begin{equation}\label{A_0_Gaussian}
{\bf A}_0({\bf r})=-\exp(-{\bf r}\cdot \hat S {\bf r})[\hat I \mbox{Tr}\hat S -\hat S]^{-1}{\bf\Omega}.
\end{equation}
As the result of substitution into equation (\ref{macro}), an ordinary differential equation for the uniform
vorticity is obtained,
\begin{equation}
\dot{\bf M}=\Big[\Big\{[\hat I \mbox{Tr}\hat S -\hat S]^{-1}({\bf M}-2{\bf\Omega})
+\frac{\Gamma\tilde\Lambda}{2\pi}\frac{{\bf M}}{|{\bf M}|}\Big\}\times \hat S{\bf M}\Big].
\end{equation}
Two integrals of motion follow from the structure of this equation (it is implied that vector ${\bf\Omega}$ 
is time-independent):
\begin{eqnarray}
{\bf M}\cdot\hat S{\bf M}=J_1,&&\\
({\bf M}-2{\bf\Omega})\cdot [\hat I \mbox{Tr}\hat S -\hat S]^{-1}({\bf M}-2{\bf\Omega})
+\frac{\Gamma\tilde\Lambda}{\pi}|{\bf M}|=J_2.&&
\end{eqnarray}
Phase trajectories of vector  ${\bf M}$ are lines of intersection between the corresponding families 
of 2D surfaces in 3D space. Depending on values of the parameters appearing in
the above expressions, phase portraits can be qualitatively different. Several examples are shown in Figs.1-2
in dimensionless quantities, where the scale of length is $R_\perp$, while the scale of inverse time is
the combination $\Gamma/\pi R_\perp^2$.

\begin{figure}
\begin{center}
 \epsfig{file=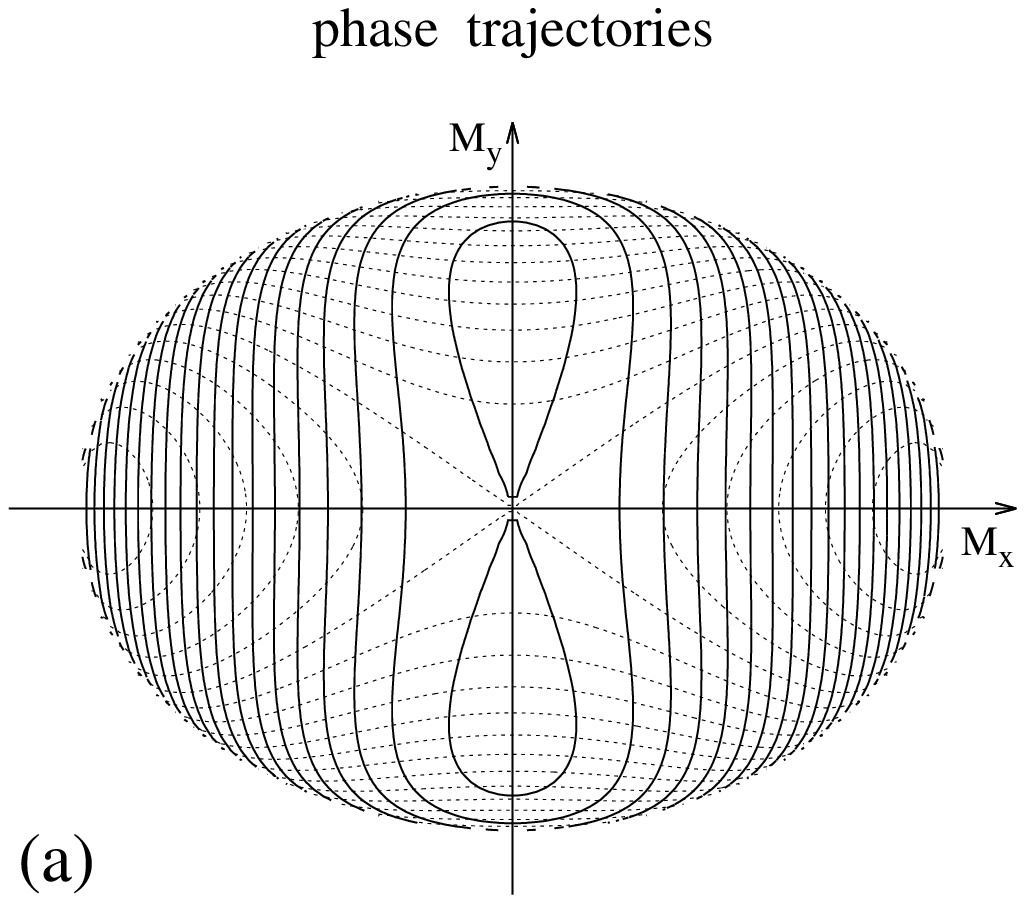, width=75mm}\\
 \epsfig{file=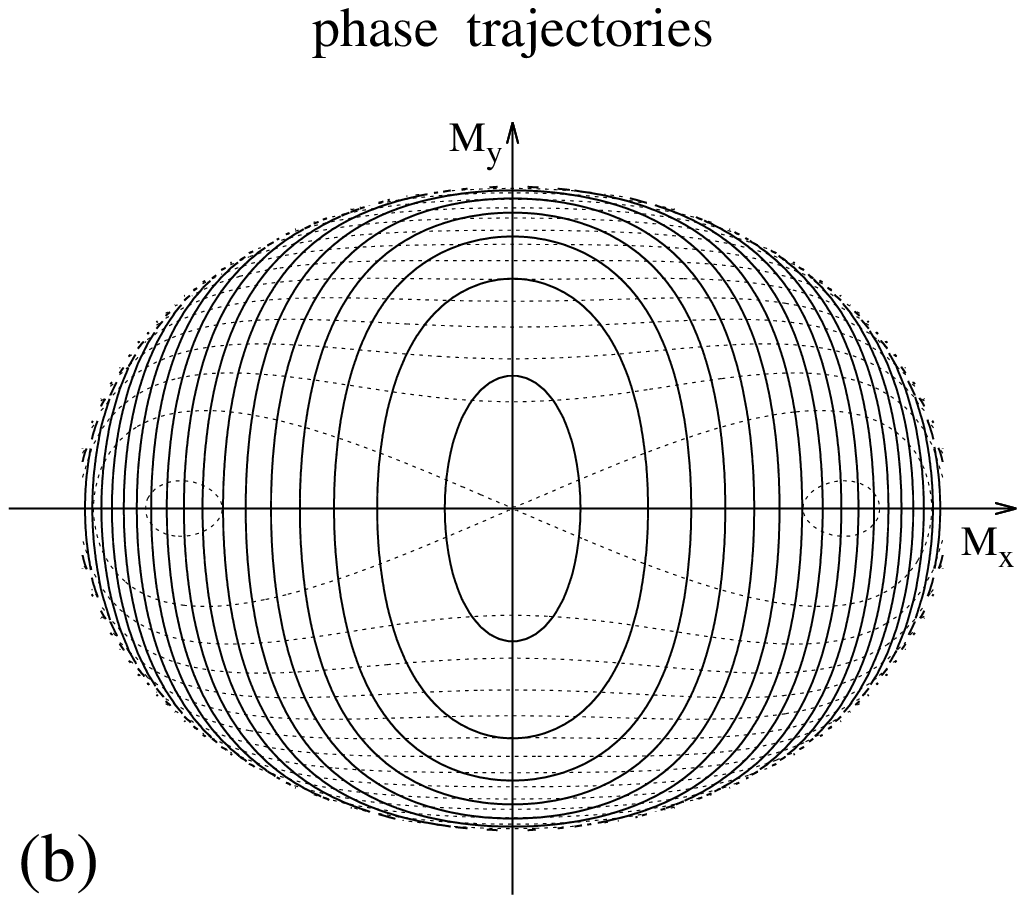, width=75mm}\\
 \epsfig{file=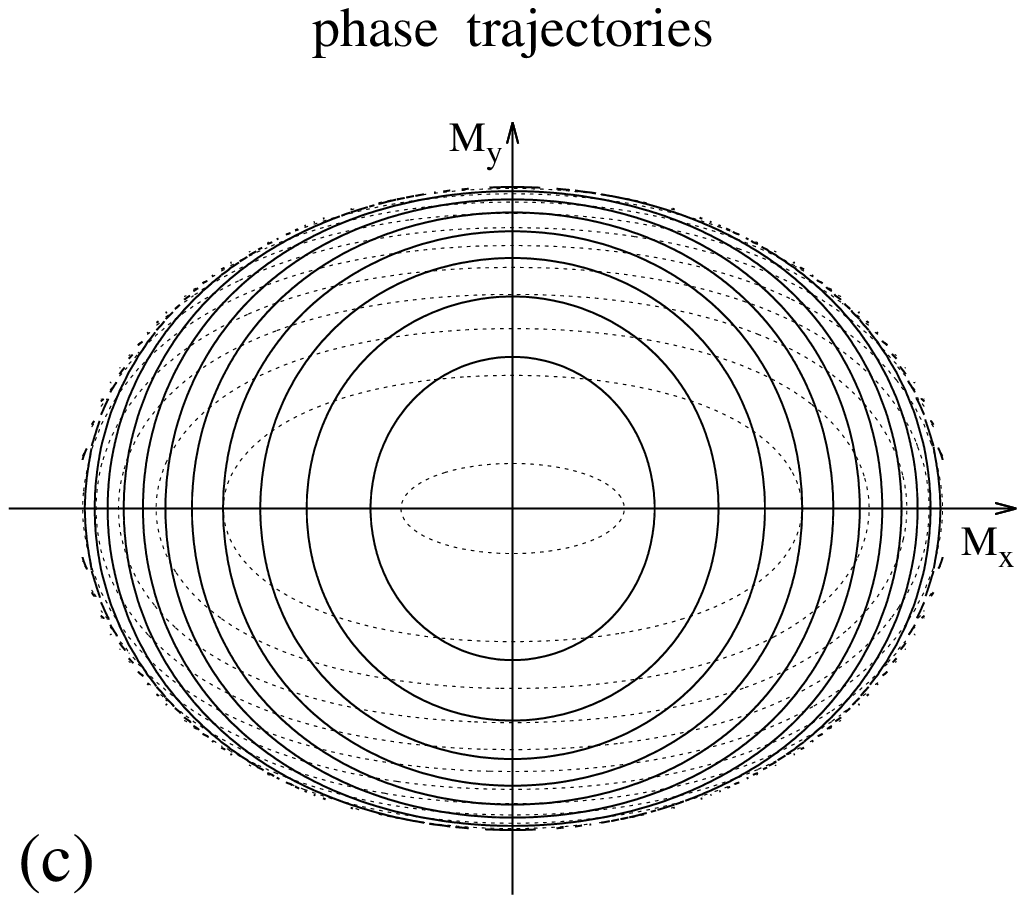, width=75mm}
\end{center}
\caption{Phase trajectories of vector  ${\bf M}$ on the ellipsoid  ${\bf M}\cdot\hat S{\bf M}=1600$
at $\tilde\Lambda =7.0$, $s_1=0.7$, $s_2=1.3$, $s_3=1.0$ 
(that is, the rotation occurs around the middle axis of the ellipsoid) for three values of the rotation frequency: 
a) $\Omega=2.0$, b) $\Omega=4.0$, c) $\Omega=10.0$. The view is from the direction of rotation axis $z$. 
The dotted lines correspond to the ``southern'' part of the ellipsoid.}
\label{ex1} 
\end{figure}
\begin{figure}
\begin{center}
 \epsfig{file=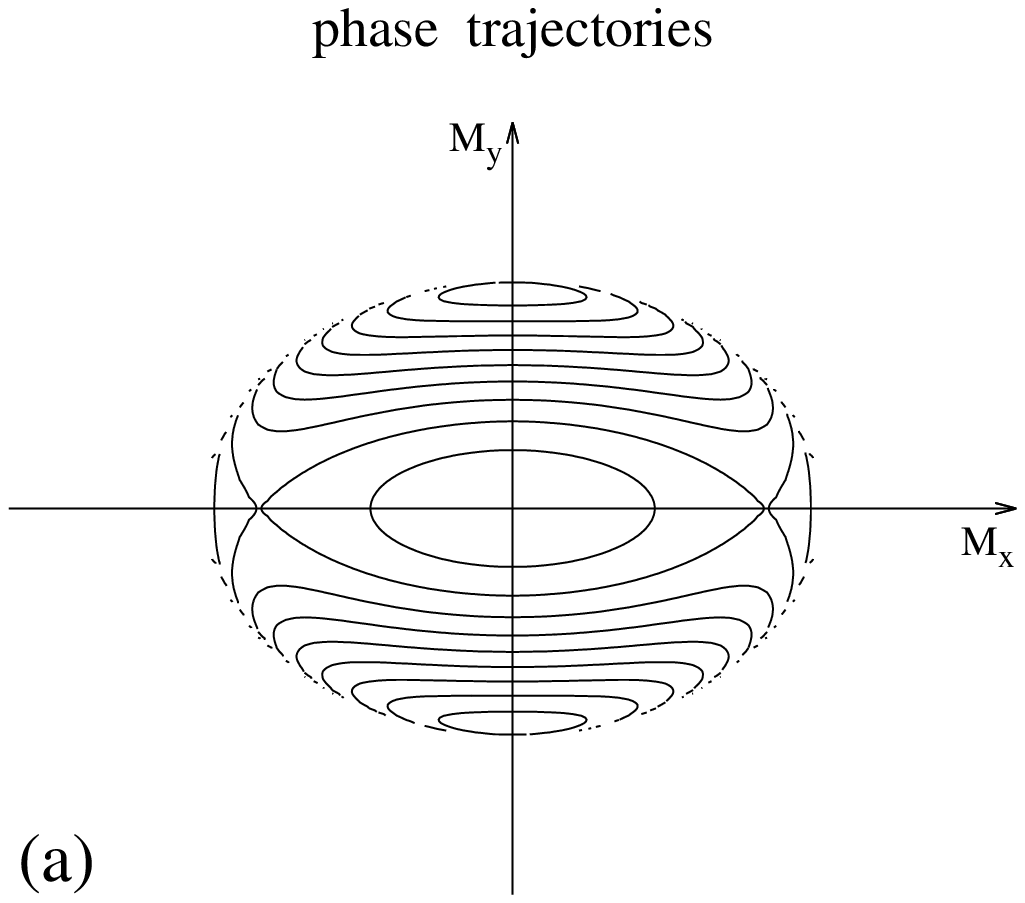, width=75mm}\\
 \epsfig{file=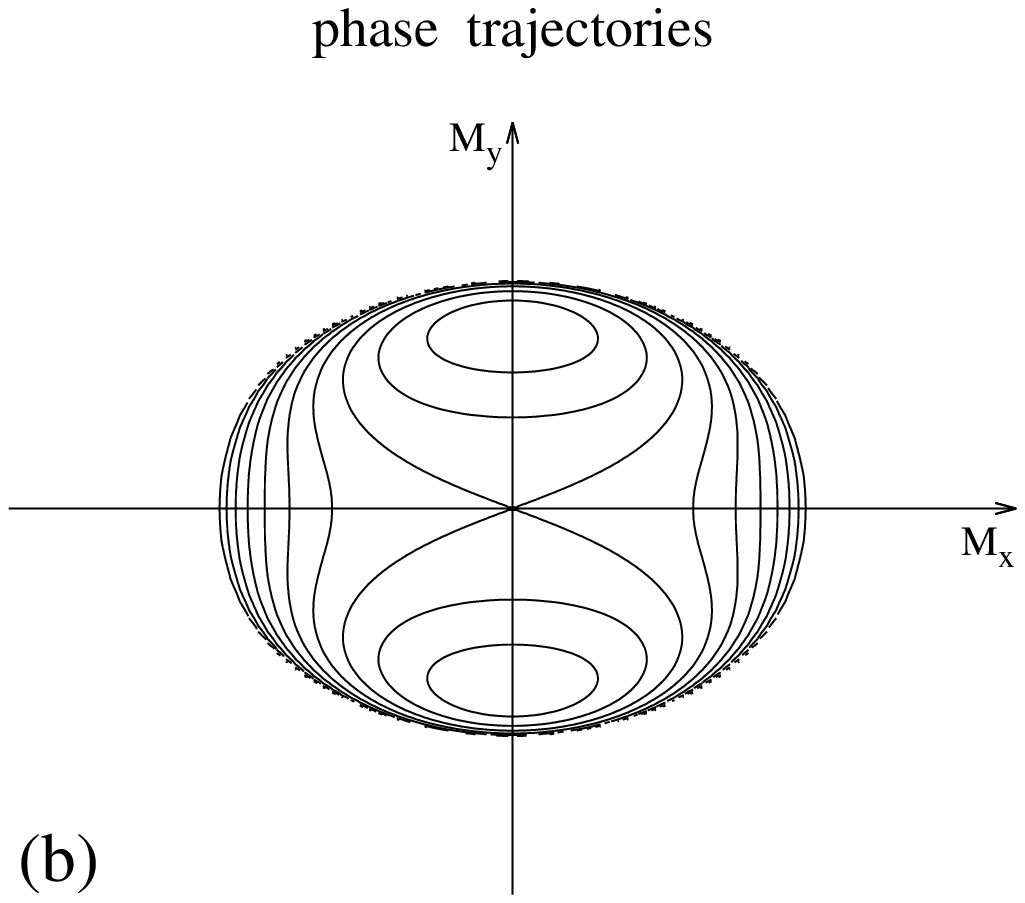, width=75mm}\\
 \epsfig{file=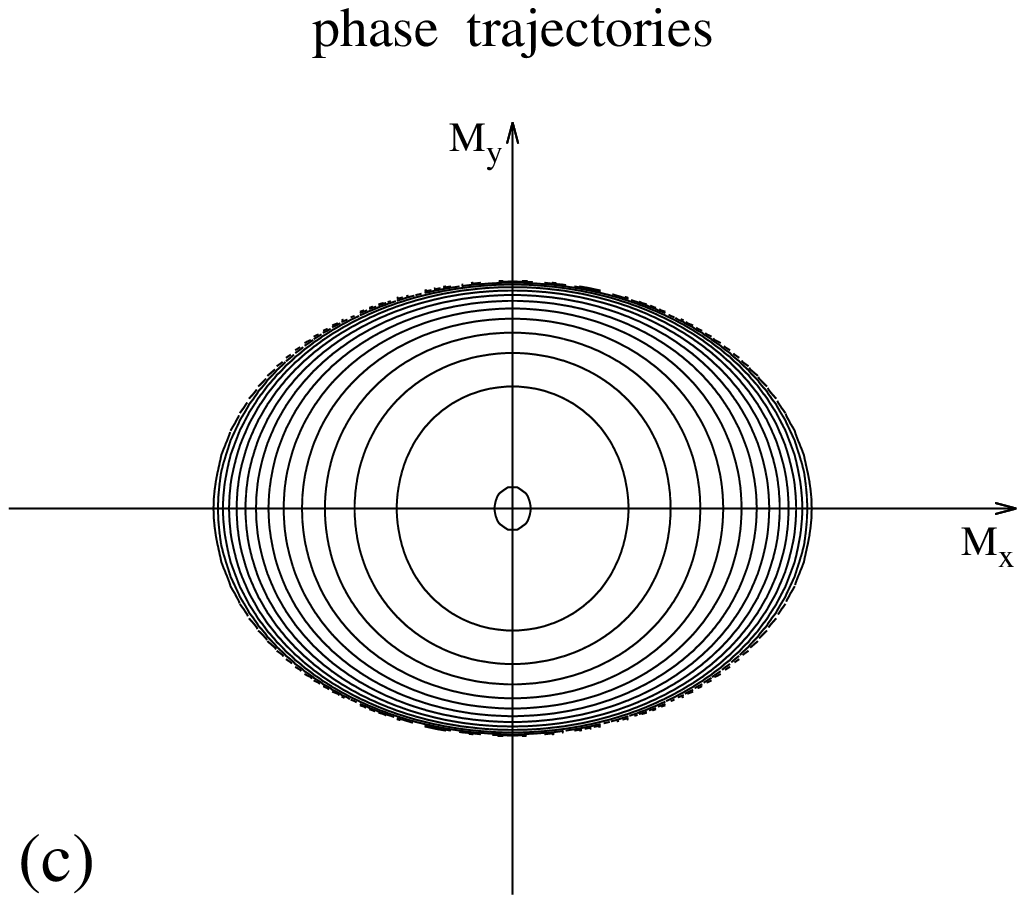, width=75mm}
\end{center}
\caption{Phase trajectories of vector  ${\bf M}$ on the ``northern'' part of the ellipsoid 
${\bf M}\cdot\hat S{\bf M}=800$ at $\tilde\Lambda =7.0$, $s_1=0.7$, $s_2=1.3$, $s_3=0.5$ 
(rotation around the large axis) for three values of the rotation frequency: 
a) $\Omega=3.0$, b) $\Omega=6.0$, c) $\Omega=12.0$. When oriented along $z$ axis, vector ${\bf M}$ 
has the same absolute value (equal to 40.0), as in Fig.1 in analogous direction.}
\label{ex2} 
\end{figure}

\section{Calculation of Green function} 

If in a condensate with Gaussian density there are only a few vortices, then for accurate description
of their interaction it is necessary to know the Green function. For that one has to solve the following 
system of linear partial differential equations, where ${\bf j}=\mbox{curl}\,{\bf A}$:
\begin{equation}
2[\hat S{\bf r}\times{\bf j}]+[\nabla\times {\bf j}]={\omegabold}({\bf r})e^{-{\bf r}\cdot \hat S {\bf r}}, 
\qquad (\nabla\cdot {\bf j})=0.
\end{equation}
Fortunately, the solution can be found by sequential use of several standard methods of mathematical physics.
Let us rewrite the given system in Fourier representation:
\begin{equation}
2[\hat S{\bf \nabla_k}\times{\bf j_k}]+[{\bf k}\times {\bf j_k}]=
-i\int \omegabold_{\bf q}e^{-{\bf r}\cdot \hat S {\bf r}+i({\bf q}-{\bf k})\cdot{\bf r}} d{\bf r}
\frac{d{\bf q}}{(2\pi)^3}, 
\end{equation}
with condition $({\bf k}\cdot {\bf j_k})=0$. Integrating over ${\bf r}$ in the r.h.s. and using the
substitution  ${\bf j_k}={\bf C_k}\exp(-{\bf k}\cdot \hat S^{-1} {\bf k}/4)$, we obtain equation
\begin{eqnarray}
&&2[\hat S{\bf \nabla_k}\times{\bf C_k}]=-i\frac{\pi^{3/2}}{\sqrt{\mbox{Det}\hat S}}\nonumber\\
&&\times\int\omegabold_{\bf q}\exp(-{\bf q}\cdot \hat S^{-1} {\bf q}/4 
+{\bf k}\cdot \hat S^{-1} {\bf q}/2)
\frac{d{\bf q}}{(2\pi)^3}. 
\end{eqnarray}
Its general solution is
\begin{eqnarray}\label{C_solution}
{\bf C_k}&=&-i\frac{\pi^{3/2}}{\sqrt{\mbox{Det}\hat S}}
\int[{\bf B_q}\exp({\bf k}\cdot \hat S^{-1} {\bf q}/2)+\hat S\nabla_{\bf k}f({\bf k},{\bf q})]\nonumber\\
&&\times \exp(-{\bf q}\cdot \hat S^{-1} {\bf q}/4)
\frac{d{\bf q}}{(2\pi)^3}, 
\end{eqnarray}
where ${\bf B_q}$  obeys condition $[{\bf q}\times{\bf B_q}]=\omegabold_{\bf q}$, and $f({\bf k},{\bf q})$ 
is an arbitrary scalar function. We choose $f$ in such a way that transversality condition 
$({\bf k}\cdot {\bf C_k})=0$ will be satisfied:
\begin{equation}
({\bf k}\cdot{\bf B_q})\exp({\bf k}\cdot \hat S^{-1} {\bf q}/2) +(\hat S{\bf k}\cdot\nabla_{\bf k})f=0.
\end{equation}
This equation is solved by the method of characteristics:
\begin{equation}
f=-\int_{-\infty}^0  ({\bf B_q}\cdot e^{\tau\hat S}{\bf k})
\exp({\bf q}\cdot \hat S^{-1}e^{\tau\hat S}{\bf k}/2)d\tau,
\end{equation}
where the matrix exponent is defined in the standard manner. It is easy to make sure with the help of 
integration by parts and application of the formula for double vector cross-product that 
\begin{eqnarray}
{\bf B_q}\exp({\bf k}\cdot \hat S^{-1} {\bf q}/2)+\hat S\nabla_{\bf k}f&&\nonumber\\
=\frac{1}{2}\int\limits_{-\infty}^0 [{\bf k}\times[e^{\tau\hat S}{\bf B}\times e^{\tau\hat S}{\bf q}]]
\exp({\bf q}\cdot \hat S^{-1}e^{\tau\hat S}{\bf k}/2) d\tau.&&
\end{eqnarray}
Now we use the equality
$$[e^{\tau\hat S}{\bf q}\times e^{\tau\hat S}{\bf B_q}]=
\exp[\tau(\hat I\mbox{Tr}\hat S -\hat S)]\omegabold_{\bf q},$$
and also we recall that
$${\bf j_k}=i[{\bf k}\times{\bf A_k}]={\bf C_k}\exp(-{\bf k}\cdot \hat S^{-1} {\bf k}/4).$$
Since, by definition,
\begin{equation}
{\bf A_k}=\int \hat G_{{\bf k},{\bf q}} \omegabold_{\bf q} \frac{d{\bf q}}{(2\pi)^3},
\end{equation}
it follows from the obtained solution for ${\bf C_k}$ that in Fourier representation the Green function is given
by expression 
\begin{eqnarray}\label{G_kq}
&&\hat G_{{\bf k},{\bf q}}=\frac{\pi^{3/2}}{2\sqrt{\mbox{Det}\hat S}}
\exp\Big[{-\frac{{\bf k}\cdot \hat S^{-1} {\bf k}}{4} -\frac{{\bf q}\cdot \hat S^{-1} {\bf q}}{4}}\Big]
\nonumber\\
&& \times\int\limits_{-\infty}^0  \exp[{{\bf q}\cdot \hat S^{-1}e^{\tau\hat S}{\bf k}/2}]
\exp[\tau(\hat I\mbox{Tr}\hat S -\hat S)]d\tau ,
\end{eqnarray}
which contains integration over an auxiliary variable $\tau$.

In order to find Green function in the physical space, it is necessary to perform the Fourier transforms on
${\bf k}$ and ${\bf q}$, with opposite signs in front of $i$ in the exponents. Having executed Gaussian 
integration, we arrive at required formula:
\begin{eqnarray}\label{G_r1_r2}
&&\hat G({\bf r}_2,{\bf r}_1)=\frac{\sqrt{\mbox{Det}\hat S}}{2\pi^{3/2}}\int\limits_{-\infty}^0 
\frac{\exp[\tau(\hat I\mbox{Tr}\hat S -\hat S)]}{\sqrt{\mbox{Det}[\hat I-e^{2\tau\hat S}]}}
\nonumber\\
&&\times\exp\Big[-{\bf r}_2\cdot \hat S[\hat I-e^{2\tau\hat S}]^{-1} {\bf r}_2-
{\bf r}_1\cdot \hat S[\hat I-e^{2\tau\hat S}]^{-1} {\bf r}_1\Big]\nonumber\\
&&\times\exp\Big[2\,{\bf r}_2\cdot \hat S e^{\tau\hat S}[\hat I-e^{2\tau\hat S}]^{-1} {\bf r}_1\Big]d\tau.
\end{eqnarray}
Though this integral cannot be expressed in elementary functions, but it can be successfully investigated
by standard analytical and numerical methods.

Now we indicate the most simple particular cases. Let the matrix $\hat S=\hat I$, so the density
distribution is spherically symmetric. Then $\hat G=\hat I {\mathbb G}({\bf r}_1,{\bf r}_2)$, where
\begin{eqnarray}
{\mathbb G}\!=\!\frac{1}{2\pi^{3/2}}\!\int\limits_0^1\!\frac{\eta d\eta}{\sqrt{(1-\eta^2)^3}}
\exp\!\Big[\frac{2\eta({\bf r}_1\cdot{\bf r}_2)\!-\!({\bf r}_1^2\!+\!{\bf r}_2^2)}{1-\eta^2}\Big]&&\nonumber\\
=\frac{1}{2\pi^{3/2}}\!\int\limits_1^\infty \!\exp\!\big[2({\bf r}_1\cdot{\bf r}_2)\chi\sqrt{\chi^2-1}
-({\bf r}_1^2+{\bf r}_2^2)\chi^2\big]d\chi&&\nonumber\\
=\frac{1}{4\pi^{3/2}}\exp\Big[-\frac{({\bf r}_1^2+{\bf r}_2^2)}{2}\Big]&&\nonumber\\
\times\!\int\limits_1^\infty\!\Big(1\!-\!\frac{1}{\upsilon^2}\Big)
\exp\Big[-\!\frac{({\bf r}_1\!-\!{\bf r}_2)^2}{4} \upsilon^2 
-\frac{({\bf r}_1\!+\!{\bf r}_2)^2}{4\upsilon^2}\Big]d\upsilon.&&
\end{eqnarray}
It is easy to check that at small difference  $|{\bf r}_1-{\bf r}_2|$ this function has ``correct'' asymptotics.

The second particular case, which will be considered in more detail, is a system of straight vortex filaments
parallel to the symmetry axis of a rotationally symmetric ellipsoid [that is, $\hat S=\mbox{Diag}(1,1,\lambda)$].
Then the velocity field turns out to be exactly two-dimensional, and interaction between point vortices in a 
transverse plane is described by a scalar Green function ${\mathsf G}({\bf x}_1,{\bf x}_2)$ that depends upon a 
pair of 2D vectors. Some mathematically equivalent integral representations for ${\mathsf G}$ are 
given below:
\begin{eqnarray}
{\mathsf G}=\frac{1}{2\pi}\!\int\limits_0^1\!\frac{\eta d\eta}{(1-\eta^2)}
\exp\!\Big[\frac{2\eta({\bf x}_1\cdot{\bf x}_2)-({\bf x}_1^2+{\bf x}_2^2)}{1-\eta^2}\Big]&&\nonumber\\
=\frac{1}{4\pi}\int\limits_1^\infty \exp\big[2({\bf x}_1\cdot{\bf x}_2)\sqrt{\zeta^2-\zeta}
-({\bf x}_1^2+{\bf x}_2^2)\zeta\big]\frac{d\zeta}{\zeta}&&\nonumber\\
=\frac{1}{4\pi}\exp\Big[-\frac{({\bf x}_1^2+{\bf x}_2^2)}{2}\Big]&&\nonumber\\
\times\!\int\limits_1^\infty\!  \frac{(u-1)}{u(u+1)}
\exp\!\Big[-\frac{({\bf x}_1\!-\!{\bf x}_2)^2}{4} u -\frac{({\bf x}_1\!+\!{\bf x}_2)^2}{4u}\Big]du.&&
\label{G_2D}
\end{eqnarray}
The last integral will be used in the next section for evaluating a theory that by different reasons suggests 
an approximate expression for Green function in a more general 2D case. Let us also note that from expression 
(\ref{G_2D}) it is easy to obtain the main asymptotics at small differences
$|{\bf x}_1-{\bf x}_2|$:
\begin{equation}
{\mathsf G}\approx \frac{1}{2\pi}\exp\Big[-\frac{({\bf x}_1^2+{\bf x}_2^2)}{2}\Big]\log(|{\bf x}_1-{\bf x}_2|^{-1}).
\end{equation}

\section{The Hamiltonian of 2D vortex dynamics}

Although equations of motion with spatially nonuniform corrections for systems of point vortices in Bose-Einstein
condensates were suggested in work  \cite{SR2004},  their Hamiltonian formulation was absent up to the moment.
Moreover, an attempt to construct a canonical theory similarly to that for a uniform fluid has resulted in a rather
rough model which sacrificed nonuniform effects in calculation of pair interactions between vortices \cite{v1,v2,v3}.
Inhomogeneity appeared there only in the form of a locally induced vortex velocity proportional to density gradient.
Here we suggest a more accurate {\it noncanonical} Hamiltonian theory for condensates with density profiles as
$\rho_0=Z(z) P(q)$, where $q=([1+\epsilon]x^2+[1-\epsilon]y^2)$. Though such a restriction on the transversal 
factor is not fundamental, but with this choice the contribution of rotation to equations of motion 
takes a particularly simple form \cite{R2016}.

Let the motion of $n$-th vortex be described by a pair of functions ${\bf x}_n(t) = (x_n(t), y_n(t))$.
Since in the l.h.s. of equation (\ref{var_eq}) a non-uniform density appears as a factor, the 2D
Cartesian coordinates of vortices cannot serve as canonical variables. A required refinement is that the dynamics 
of the system is determined by noncanonical Hamiltonian equations
\begin{equation}
\sigma_n P(q_n)\dot  x_n= \frac{\partial H}{\partial y_n},\qquad 
-\sigma_n P(q_n)\dot  y_n=\frac{\partial H}{\partial x_n},
\end{equation}
where $\sigma_n=\pm 1$ depending on positive or negative vortex orientation, 
$q_n=[1+\epsilon]x_n^2+[1-\epsilon]y_n^2$. The Hamiltonian  $H$ is a sum of contributions by the local induction,
by rotation, and also by all pair interactions (for convenience, all the quantities are non-dimensionalized):
\begin{eqnarray}\label{H_pv}
&& H=\sum_n \Big[\frac{\tilde\Lambda}{2}P(q_n)
-\frac{\tilde\Omega\sigma_n}{2}\int\limits_{q_n}^{\infty}P(q)dq\Big]\nonumber\\
&&\qquad+\sum_{m\neq n}\frac{\sigma_n\sigma_m}{2}  G({\bf x}_n,{\bf x}_m),
\end{eqnarray}
where $\tilde \Omega$ is a dimensionless frequency of rotation, 
$G({\bf x}_n,{\bf x}_m)$ is the corresponding Green function (multiplied by  $2\pi$).
The equations of motion then take form
\begin{eqnarray}\label{xn_t}
\dot  x_n&=&\Big\{ \tilde\Lambda\sigma_n \frac{P'(q_n)}{P(q_n)} +\tilde\Omega\Big\}[1-\epsilon]y_n\nonumber\\
&&+\frac{1}{P(q_n)}\sum_{m\neq n}\sigma_m\frac{\partial G({\bf x}_n,{\bf x}_m)}{\partial y_n},
\end{eqnarray}
\begin{eqnarray}\label{yn_t}
-\dot  y_n&=&\Big\{ \tilde\Lambda\sigma_n \frac{P'(q_n)}{P(q_n)} +\tilde\Omega\Big\}[1+\epsilon]x_n\nonumber\\
&&+\frac{1}{P(q_n)}\sum_{m\neq n}\sigma_m\frac{\partial G({\bf x}_n,{\bf x}_m)}{\partial x_n}.
\end{eqnarray}
Note that in Gaussian case the expression in curly brackets in equations (\ref{xn_t}) and (\ref{yn_t}) does not 
depend on $x_n$ and $y_n$, since $P(q_n)=\exp(-q_n)$, and the Green function (with $\epsilon =0$) is equal to
\begin{equation}
G({\bf x}_n,{\bf x}_m)=\exp(-q_n/2)\exp(-q_m/2){\mathsf g}(a,b), 
\end{equation}
where $a={({\bf x}_n-{\bf x}_m)^2}/{4}$, $b= {({\bf x}_n+{\bf x}_m)^2}/{4}$,
\begin{equation}\label{g_exact_Gauss}
{\mathsf g}(a,b)=\frac{1}{2}\int\limits_1^\infty  
\frac{(u-1)}{u(u+1)}\exp\Big[-a u -\frac{b}{u}\Big]du.
\end{equation}

In order to suggest a reasonable approximation for $G$ at different density profiles, we first rewrite equation
(\ref{A_eq}) as follows:
\begin{equation}
-\sqrt{\rho}\nabla\cdot\Big(\frac{1}{\rho}\nabla \sqrt{\rho}\Big[\frac{A}{\sqrt{\rho}}\Big]\Big)
=\sqrt{\rho}\omega({\bf x}),
\end{equation}
where $A({\bf x})$ and $\omega({\bf x})$ are $z$-components of the corresponding vectors
(while the transversal components are equal to  zero). Form here we easily obtain that function 
 $\psi=A/\sqrt{\rho}$ satisfies the equation
\begin{equation}\label{psi_eq}
\tilde\kappa^2({\bf x})\psi  -\Delta\psi =\sqrt{\rho}\omega,
\end{equation}
where
\begin{equation}
\tilde\kappa^2({\bf x})\equiv\sqrt{\rho}\Delta\frac{1}{\sqrt{\rho}}.
\end{equation}
Consider a special class of densities $\rho_\kappa(x,y)$, for which
$\tilde\kappa^2({\bf x})=\kappa^2=\mbox{const}>0$. In such a case 
\begin{equation}
(\partial_x^2+\partial_y^2)\frac{1}{\sqrt{\rho_\kappa}}=\kappa^2\frac{1}{\sqrt{\rho_\kappa}}.
\end{equation}
The general solution of this equation is represented by formula
\begin{equation}
\rho_\kappa=\Big[\int_0^{2\pi}C(\phi)\exp(\kappa x\cos\phi+\kappa y\sin\phi)\frac{d\phi}{2\pi}\Big]^{-2},
\end{equation}
where $C(\phi)$ is an arbitrary non-negative function (in particular case $C(\phi)=1$ the integral is equal to
the modified Bessel function $I_0(\kappa|{\bf x}|)$, that corresponds to the density profile suggested in work
 \cite{RP1994}). Equation (\ref{psi_eq}) is then easily solved, and it follows from the solution that the Green
 function is given by exact formula
\begin{equation}
G_\kappa({\bf x}_1,{\bf x}_2)=\sqrt{\rho_\kappa({\bf x}_1) \rho_\kappa({\bf x}_2)}
K_0(\kappa|{\bf x}_1-{\bf x}_2|),
\end{equation}
where $K_0$ is the corresponding Bessel function.

In general case the factor $\sqrt{\rho({\bf x}_1) \rho({\bf x}_2})$ remains in expression for the Green function, 
while instead of  $K_0(\kappa|{\bf x}_1-{\bf x}_2|)$ there stands the Green function $g({\bf x}_1,{\bf x}_2)$
(also symmetric on the arguments) of spatially non-uniform operator $(\tilde\kappa^2({\bf x})  -\Delta)$.
Clearly, if $\tilde\kappa^2({\bf x})$ is positive everywhere then  $g$ has the same asymptotics at close
arguments, as $K_0$ with a local value  $\tilde\kappa$. A good approximation is
\begin{equation}
g({\bf x}_1,{\bf x}_2)\approx 
K_0\Big(\mbox{min}\int_{{\bf x}_1}^{{\bf x}_2}\tilde\kappa({\bf x})|d{\bf x}|\Big),
\end{equation}
where the minimum is taken over all possible paths connecting points ${\bf x}_1$ and ${\bf x}_2$.
But in that case, to find the minimizing path, it is necessary to solve the corresponding ordinary
differential equation of the second order. As a quite ``cheap'' variant, we can suggest to take the inhomogeneity
into account by the following replacement:
\begin{equation}\label{G_appr}
G({\bf x}_1,{\bf x}_2)\approx \sqrt{\rho({\bf x}_1) \rho({\bf x}_2})
K_0\Big(\sqrt{\frac{\tilde\kappa^2_1 +\tilde\kappa^2_2}{2}}|{\bf x}_1-{\bf x}_2|\Big),
\end{equation}
where  $\tilde\kappa_{1,2}=\tilde\kappa({\bf x}_{1,2})$. The question ``what to do if  $\tilde\kappa^2({\bf x})$
takes negative values in some domain?'' remains open.

\begin{figure}
\begin{center}
 \epsfig{file=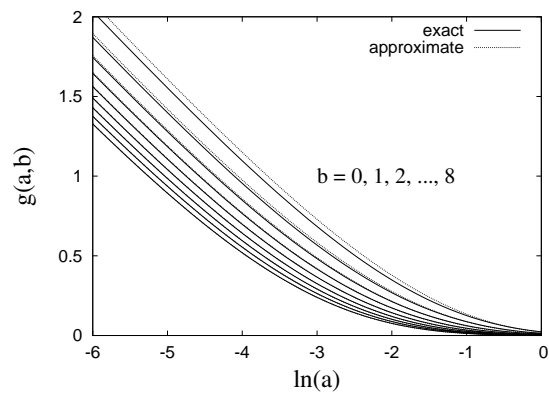, width=75mm}
\end{center}
\caption{Numerical comparison of formulas (\ref{g_exact_Gauss}) and (\ref{g_appr_Gauss}).}
\label{comparison} 
\end{figure}

It is possible to  evaluate quality of approximation (\ref{G_appr}) using Gaussian density as an example
and comparing the exact integral  (\ref{g_exact_Gauss}) with the corresponding approximation
\begin{equation}\label{g_appr_Gauss}
{\mathsf g}_{\rm appr}(a,b)=K_0\big(2\sqrt{a}\sqrt{2+a+b}\big).
\end{equation}
Such a comparison is shown in Fig.3. It is seen the approximate theory  works quite well in this case 
up to values $a\lesssim 1$.

It is necessary to note about an ``inequality of rights'' taking place for vortices situated at points with
different density  $\rho$. Since the Green function contains factor $\sqrt{\rho({\bf x}_n) \rho({\bf x}_m})$,
and the division by $\rho({\bf x}_n)$ occurs when equations of motion are compiled, therefore in principal terms 
the ratio $\sqrt{\rho({\bf x}_m) /\rho({\bf x}_n})$ appears which results in an hierarchy of one vortex influence 
upon another. A vortex which is more close to the trap center (where the density is larger) acts rather strongly
upon a peripheral vortex, while the last one practically is unable to influence upon the central vortex.
Therefore the dynamics of one or several ``nearly equal in rank'' central vortices can turn out ot be 
quasi-autonomous, at the same time essentially determining the motion of all peripheral vortices.

As an example of application of approximate formula (\ref{g_appr_Gauss}), we now consider an integrable system
of two vortices at $\epsilon=0$, which has two integrals of motion --- the Hamiltonian and the angular momentum
[the term in (\ref{H_pv}) proportional to  $\tilde\Omega$]:
\begin{eqnarray}
\frac{\tilde\Lambda}{2}[P_1+P_2]+\sigma_1 \sigma_2 \sqrt{P_1 P_2}
 K_0\Big(\Big[2-\frac{1}{2}\log(P_1 P_2)\Big]^{\frac{1}{2}}&&
\nonumber\\
\times\Big[\!-\!\log(P_1 P_2)\!-\!2\sqrt{\log(P_1)\log(P_2)}\cos\phi\Big]^{\frac{1}{2}}\Big)\!=\!h_0,&&
\end{eqnarray}
\begin{equation}
\sigma_1 P_1+ \sigma_2 P_2=2M,
\end{equation}
where  $\phi$ is the angle between vectors ${\bf x}_1$ and ${\bf x}_2$. It is convenient to introduce new variables
$$
p_+=(P_1+P_2)/2,\qquad  p_-=(P_1-P_2)/2.
$$
Let for definiteness $\sigma_1=+1$. Then, depending on the sign  $\sigma_2=\sigma$, we have two possible variants:
either $p_+=M$ with $\sigma=+1$, or $p_-=M$ with $\sigma=-1$. In the new variables the equation 
\begin{eqnarray}
&&\tilde\Lambda p_+ + \sigma \sqrt{(p_+^2- p_-^2)}
 K_0\Big(\Big[2-\frac{1}{2}\log(p_+^2- p_-^2)\Big]^{\frac{1}{2}}
\nonumber\\
&&\times\Big\{-\log(p_+^2- p_-^2)\nonumber\\
&&-2\sqrt{\log(p_+ + p_-)\log(p_+ - p_-)}\cos\phi\Big\}^{\frac{1}{2}}\Big)=h_0
\end{eqnarray}
determines phase trajectories of the system either in the plane $(p_-,\phi)$  at $\sigma=+1$, $p_+=M$, or
in the plane $(p_+,\phi)$ at  $\sigma=-1$, $p_-=M$.

Finally, we indicate one else interesting application of the approximate formula (\ref{G_appr}). 
Let the condensate density be independent on $z$ coordinate. Consider a long-wave dynamics of several vortex
filaments that is described by unknown functions $x_n(z,t)$ and $y_n(z,t)$ slowly depending on  $z$. 
Then equations of motion take the form
\begin{equation}
 \sigma_n P(q_n)\partial_t  x_n= \frac{\delta H_*}{\delta y_n},\quad 
-\sigma_n P(q_n)\partial_t  y_n= \frac{\delta H_*}{\delta x_n},
\end{equation}
where $H_*=\int h dz$, and the line density of Hamiltonian is equal to 
\begin{eqnarray}\label{h}
h=\sum_n \Big[\frac{\tilde\Lambda}{2}P(q_n)\Big[1+\frac{(x'^2_n+y'^2_n)}{2}\Big]
-\frac{\tilde\Omega\sigma_n}{2}\int\limits_{q_n}^{\infty}P(q)dq\Big]&&
\nonumber\\
+\!\sum_{m\neq n}\!\frac{\sigma_n\sigma_m}{2} \sqrt{P(q_n)P(q_m)} 
K_0\Big(\!\sqrt{\frac{\tilde\kappa^2_n\! +\!\tilde\kappa^2_m}{2}}|{\bf x}_n\!-\!{\bf x}_m|\!\Big).&&
\end{eqnarray}
Herewith
\begin{eqnarray}
\tilde\kappa^2(x,y)=-2\frac{P'(q)}{P(q)}\qquad\qquad \qquad\qquad \qquad\qquad &&\nonumber\\
+\Big(3\frac{P'^2(q)}{P^2(q)}-2\frac{P''(q)}{P(q)}\Big)
[(1+\epsilon)^2x^2+(1-\epsilon)^2y^2].&&
\end{eqnarray}
In the Gaussian density case
$$
\tilde\kappa^2_n=2+(1+\epsilon)^2x_n^2+(1-\epsilon)^2y_n^2.
$$
Relatively compact expression for the line density of Hamiltonian is at $\epsilon=0$, $\tilde\Omega=0$
in terms of complex functions  $w_n(z,t)=x_n(z,t)+iy_n(z,t)$:
\begin{eqnarray}\label{h_w}
&&h_w=\sum_n \Big[\frac{\tilde\Lambda}{2}e^{-w_n w^*_n}\Big(1+\frac{w'_n {w^*_n}'}{2}\Big)\Big]
\nonumber\\
&&\quad+\sum_{m\neq n}\frac{\sigma_n\sigma_m}{2} \exp(-[{w_n w_n^* +w_m w_m^*}]/{2})\nonumber\\
&&\qquad\times K_0\big(|w_n-w_m|\sqrt{2+(|w_n|^2+|w_m|^2)/2}\big).
\end{eqnarray}
The corresponding equations of motion are
\begin{eqnarray}\label{w_n_t}
&&i\partial_t w_n=\sigma_n\tilde\Lambda\Big[-\frac{1}{2}w_n''-w_n+\frac{{w'_n}^2w_n^*}{2}\Big]\nonumber\\
&&\qquad\quad-w_n\sum_{m\neq n}\sigma_m \exp[(|w_n|^2  -|w_m|^2)/2]\nonumber\\
&&\qquad\qquad\times K_0\big(|w_n-w_m|\sqrt{2+(|w_n|^2+|w_m|^2)/2}\big)\nonumber\\
&&\qquad\quad+\sum_{m\neq n}\sigma_m \exp[(|w_n|^2  -|w_m|^2)/2]\nonumber\\
&&\qquad\qquad\quad\times K_0'\big(|w_n-w_m|\sqrt{2+(|w_n|^2+|w_m|^2)/2}\big)\nonumber\\
&&\qquad\qquad\qquad\times\Big\{\frac{(w_n-w_m)}{|w_n-w_m|}\sqrt{2+(|w_n|^2+|w_m|^2)/2}\nonumber\\
&&\qquad\qquad\quad\quad\quad+\frac{|w_n-w_m|w_n}{2\sqrt{2+(|w_n|^2+|w_m|^2)/2}}
\Big\}.
\end{eqnarray}
In particular, if there are only two vortex filaments of the same sign in the condensate, 
then symmetric solutions $w_1=w(z,t)$, $w_2=-w(z,t)$ are possible. For that case the following evolutionary 
equation is valid, 
\begin{eqnarray}\label{w_symm_t}
&&i w_t=\tilde\Lambda\Big[-\frac{1}{2}w''-w+\frac{{w'}^2w^*}{2}\Big]
-w K_0\big(2|w|\sqrt{2+|w|^2}\big)\nonumber\\
&&\qquad\quad+ 2K_0'\big(2|w|\sqrt{2+|w|^2}\big)\frac{(1+|w|^2)w}{|w|\sqrt{2+|w|^2}}.
\end{eqnarray}

\section {Conclusions}

Thus, the present work has demonstrated that Hamiltonian formalism is a natural and efficient tool for
theoretical study of quantum vortex dynamics in spatially nonuniform systems obeying the Gross-Pitaevskii
equation (perhaps, with a different nonlinearity; and then our results can be almost literally applied 
to superfluid Fermi-gases \cite{s_f_Fermi_gas-1,s_f_Fermi_gas-2,s_f_Fermi_gas-3}).
This approach has allowed us to obtain without extreme efforts quite nontrivial and accurate new results
in the field of quantum gases. One should not exclude also possibility of its applications in other fields
of physics, for example in astrophysics for investigation of vortical motion of matter in sufficiently
dense quasi-equilibrium gravitating systems, provided hydrodynamic effects dominate over kinetic ones.
If not always quantitative results, but a qualitative understanding of the mechanics of nonuniform liquid
systems can be given by the Hamiltonian method in this situation as well as in other ones.

Developing further the results obtained in this work, it will be possible to study still many interesting 
properties of vortices in trapped Bose-Einstein condensates. For instance, our macroscopic theory does not answer 
the question: how are the vortices in a nonuniform lattice packed --- does mainly a deformation of a perfect lattice 
take place (as it was assumed in work \cite{SR2004}), or do many defects appear? This question can be clarified
within suggested here Hamiltonian theory of point vortices through a systematic finding and investigation of
local minima of the Hamiltonian with  approximate Green function for various density profiles and for 
various numbers of vortices. Another interesting problem --- how do individual vortices move in macroscopically
quasi-stationary states in the cases when a number of vortices does not correspond to the rotational frequency
of an anisotropic trap? Do wanders from periphery to the center and back occur, or each vortex remains 
on a more-less definite orbit? One else interesting problem is: how do the obtained here 3D uniform solutions 
$\tilde{\omegabold}={\bf M}(t)$ of macroscopic equations correspond  to the dynamics of systems with a
relatively small number of vortex filaments (of order 10-50)?

In the hydrodynamic theory itself there remain still many unsolved problems concerning spatially 
inhomogeneous systems. So, to the best author's knowledge, in the 3D case the Gaussian is still, in essence,
the only density profile [not to speak about less interesting exponent $\rho_0=\exp({\bf k}_0\cdot{\bf r})$],
for which the Green function determining the vortex Hamiltonian has been calculated.
So, it is necessary to search for different exactly solvable profiles, as well as to develop efficient methods
for approximate calculation of 3D Green functions.

\end{document}